\documentclass[a4paper,11pt]{article}
\usepackage{pos}

\title{Inflaton Dark Matter}

\author*[a]{Jong-Hyun Yoon}

\affiliation[a]{Department of Physics, University of Helsinki,\\
Gustaf H\"allstr\"omin katu 2a, FI-00014 Helsinki, Finland}

\emailAdd{jong-hyun.yoon@helsinki.fi}

\abstract{We discuss a minimal extension of the Standard Model (SM) where a single real scalar field serves as both inflaton and dark matter. The corresponding Lagrangian contains the renormalizable interactions of the inflaton field. Quantum effects generally induce a non--minimal coupling to gravity which facilitates inflation consistent with the PLANCK constraints. A large fraction of the inflaton quanta produced after inflation must be converted into the SM radiation reheating the Universe and the rest remains dark matter today. We consider thermal and non--thermal production of inflaton dark matter. In the non--thermal case, we take into account collective effects with the help of lattice simulations. Combining analytic and numerical results with the unitarity consideration, we find that the inflaton dark matter model is viable only in the thermal case where the inflaton mass is near half the Higgs mass.

}

\FullConference{%
  7th Symposium on Prospects in the Physics of Discrete Symmetries (DISCRETE 2020-2021)\\
  29th November - 3rd December 2021\\
 Bergen, Norway}

\begin{document}
\maketitle

\section{Introduction}
Dark matter (DM) is one of the deepest mysteries in the history of physics, yet no convincing signatures of dark matter have been detected. Motivated by the inflationary Big Bang cosmology, one may consider an economical possibility that a single scalar field plays the role of both the inflaton field and dark matter \cite{Liddle:2008bm}. Non--minimal couplings to gravity induced by the quantum effects \cite{Chernikov:1968zm} may be used to fit the inflationary dynamics while the renormalizable interactions with the Higgs explain the reheating of the Universe. The simplest thermal DM model based on a non--minimal scalar coupling to curvature was studied in \cite{Lerner:2009xg}. The non--thermal case often requires one to perform lattice simulations \cite{Kofman:1994rk, Kofman:1997yn, Greene:1997fu, Khlebnikov:1996wr, Prokopec:1996rr}. Taking into account the unitarity of the system, we discuss the viability of the inflaton dark matter model and find the allowed parameter space \cite{Lebedev:2021zdh}.

\section{Singlet--driven inflation}
Let us study general renormalizable interactions of the real singlet scalar inflaton field $\phi$ and the Higgs field $H$ \cite{Lebedev:2011aq, Lebedev:2021xey}
 \begin{equation}
{\cal L}_{J} = \sqrt{-\hat g} \left(   -{1\over 2}  \Omega  \hat R \,  
 +  {1\over 2 } \, \partial_\mu \phi \partial^\mu \phi +  \,  (D_\mu H)^\dagger D^\mu  H  - {V(\phi,H)  }\right) \;,
\label{L-J}
\end{equation}
where $\hat g^{\mu \nu}$ is the Jordan frame metric and $\hat R$ is the corresponding scalar curvature. 
In the unitary gauge,
\begin{equation}
H(x)= {1\over \sqrt{2}} \left(
\begin{matrix}
0\\
h(x)
\end{matrix}
\right)\;,
\end{equation}
the $Z_2$--symmetric potential has been assumed to stabilize $\phi$
\begin{equation}
V(\phi, h) = {1\over 4} \lambda_h h^4 + {1\over 4} \lambda_{\phi h }h^2 \phi^2 + {1\over 4} \lambda_\phi \phi^4  +
{1\over 2} m_h^2 h^2 + {1\over 2} m_\phi^2 \phi^2   \;.
\label{potential}
\end{equation}
In Planck units ($M_{\rm Pl}=1$), the function $\Omega$ at the lowest order of non--minimal scalar--gravity couplings is given by
\begin{equation}
 \Omega = 1+  \xi_h h^2 + \xi_\phi \phi^2 \;.
 \label{Omega}
 \end{equation}
The non--minimal coupling to gravity term can be eliminated by a conformal transformation $g_{\mu\nu} = \Omega \, \hat g_{\mu\nu}  $ from the Jordan frame to the Einstein Frame.
The Einstein frame Lagrangian reads
 \begin{equation}
 {\cal L} = {3\over 4} \Bigl(\partial_\mu \ln ( \xi_h h^2 + \xi_\phi \phi^2 )\Bigr)^2 + {1\over 2} {1\over  \xi_h h^2 + \xi_\phi \phi^2} \Bigl(  (\partial_\mu h)^2 + (\partial_\mu \phi)^2 \Bigr)
 - {V \over ( \xi_h h^2 + \xi_\phi \phi^2)^2} \;.
 \end{equation} 
  Introducing the variables and normalizing them canonically
 \begin{eqnarray}
 \chi= \sqrt{3\over 2} \ln ( \xi_h h^2 + \xi_\phi \phi^2) ~~,~~   \tau = {h \over \phi} \;, ~~~~ \chi^\prime = {\chi \; \sqrt{1+{1\over 6\xi_\phi} }}~~,~~ \tau^\prime={\tau \over \sqrt{\xi_\phi}} \;,  
\end{eqnarray}
one may obtain the potential of the new inflaton field $\chi'$
 \begin{equation}
V_E = {\lambda_{\phi} \over 4 \xi_\phi^2} \left( 1+  \exp \left(     - {2 \gamma \chi^\prime \over \sqrt{6}}  \right) \right)^{-2} \;,
  \label{VE1}
 \end{equation}
 where $ \gamma = \sqrt{6 \xi_\phi \over 6\xi_\phi +1  } \;$.
The slow roll parameters are expressed in terms of the new inflaton field $\chi'$,
\begin{eqnarray}
    \epsilon = {1\over 2} \left( {\partial V_E / \partial \chi^\prime \over V_E }    \right)^2   ~~,~~  \eta = {\partial^2 V_E / \partial \chi^{\prime\,2 } \over V_E }   \;.
\end{eqnarray}
Since inflation ends when $\epsilon \simeq 1$, it determines $\chi^\prime_{\rm end}$. The number of $e$--folds is given by
 \begin{equation}
N= \int_{\rm in}^{\rm end } H \, dt = - \int^{\rm end }_{\rm in} {V_E \over \partial V_E / \partial \chi' }\, d \chi' \;.
 \end{equation}
That is, a fixed $N$ determines the initial $\chi^\prime_{\rm in}$. The COBE constraint on inflationary perturbations requires $V_E/ \epsilon \simeq 0.027^4$ at $\chi^\prime_{\rm in}$. Therefore, one finds
\begin{equation}
 {\lambda_{\phi} \over 4 \xi_\phi^2} = 4\times 10^{-7} \, {1\over \gamma^2 N^2 } \;.
 \label{cobe}
 \end{equation}
The spectral index $n$ and the tensor to scalar ratio $r$ are
   \begin{eqnarray}
     n = 1-6 \epsilon + 2 \eta \simeq 1-{2\over N}  - {9\over 2 \gamma^2 N^2 }  ~~~~,~~~~   r =16 \epsilon \simeq {12\over \gamma^2 N^2}  \;.
\label{n-r}
      \end{eqnarray} 
Our set--up $\gamma \sim 1$ fits well the PLANCK observation ($n \sim 1 $ and $r \ll 1$) for $N =50$ to 60 \cite{Akrami:2018odb}.

Since a non--minimal coupling term corresponds to a non--renormalizable dim--5
operator, the system is meaningful up to a cutoff scale, $ \Lambda \sim 1/\xi_\phi$. That is, the inflationary scale
$(\lambda_\phi/ 4\xi_\phi^2)^{1/4}$ must be below the cutoff. Combined with \eqref{cobe}, this requires at the inflationary scale
\begin{equation}
 \lambda_{\phi} (H)< 4\times 10^{-5} 
 \label{uni-bound}
 \end{equation}
and $\xi_\phi (H) <300$, where we have set $\gamma \sim 1$ for large $\xi_\phi$.

Our generalized framework allows one to reproduce the Higgs inflation model, where $\xi_h$ must be large such that the corresponding $\gamma \sim 1$ \cite{Bezrukov:2007ep}. However, low $\xi$ is also possible in the present model and it may lead to small deviations in the inflationary prediction \eqref{n-r}. The cutoff scale $\Lambda$ depends on the inflaton background value and therefore the Higgs inflation model survives the unitarity constraint during inflation \cite{Bezrukov:2010jz}. At the end of inflation and the beginning of reheating, however, this background becomes less important, so the unitarity issue remains in the Higgs inflation model.

\section{Non--thermal inflaton dark matter and reheating}
Since the inflaton amplitude decreases such that $\xi_\phi \phi^2, \xi^2_\phi \phi^2 <1 $ after inflation, one finds
\begin{equation}
\chi \simeq \phi ~~,~~ V_E (\phi) \simeq {1\over 4 } \lambda_\phi \,\phi^4\;.
\end{equation}
Therefore, the equation of motion (EOM) for the inflaton background field is
 \begin{equation} \label{eom}
 \ddot\phi + 3H \dot \phi + \lambda_\phi\, \phi^3 =0 \;.
 \end{equation}
 After a few oscillations, the solution is approximated by a Jacobi cosine function,
 \begin{equation} \label{jacobi}
\phi(t) = {\Phi_0 \over a(t)} \; {\rm cn} \left( x, {1\over \sqrt{2}} \right) \;,
 \end{equation}
where $\Phi_0$ is the initial amplitude of the inflaton background field and $x=(48 \lambda_\phi)^{1/4} \sqrt{t} $ is the conformal time.
The EOM for the Higgs in momentum space takes the form of the Lam\'e equation,
    \begin{equation}
X_k^{\prime\prime} + \left[     \kappa^2  + {\lambda_{\phi h} \over 2 \lambda_\phi} \;  {\rm cn}^2 \left( x, {1\over \sqrt{2}} \right)  \right] \,X_k=0 \;,
 \label{Lame}
 \end{equation}
where $X_k(t) \equiv a(t) h_k(t) $, $\kappa^2  \equiv {k^2 / (\lambda_\phi \Phi_0^2)}$, and the prime denotes differentiation with respect to $x$.

\begin{figure}[t] 
\centering{
\includegraphics[scale=0.34]{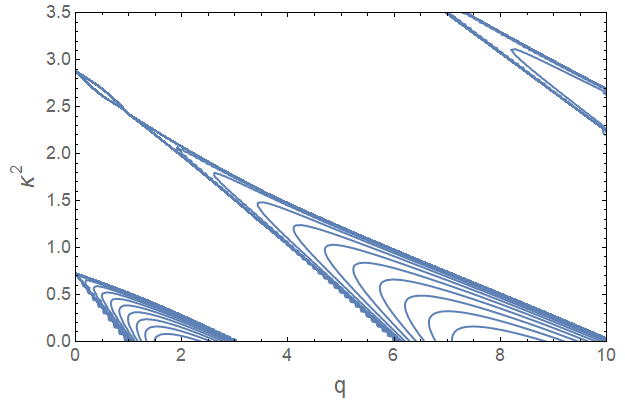}
\includegraphics[scale=0.38]{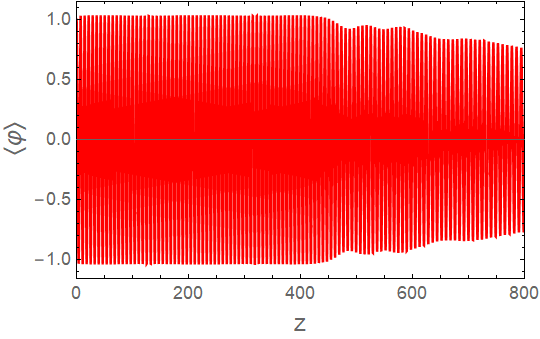}
}
\caption{ \label{Lame-fig}
{\it Left}: Instability bands of the Lam\'e  equation  with $q=  \lambda_{\phi h} / (2 \lambda_\phi)$. The Floquet exponent is constant along the contours and the innermost corresponds to the strongest resonance. {\it Right}: Decay of the classical inflaton background  
$\langle \varphi \rangle $  in the comoving frame  ($ \varphi = a\, \phi$) due to rescattering of the inflaton quanta. The conformal
time $z$ is defined by $dz =\sqrt{\lambda_\phi} \,\Phi_0  \,dt/a(t) $; $\lambda_\phi=10^{-13}$ and  $\Phi_0 =1.7$ in Planck units.
 The amplitude is normalized to 1 at the initial point.
}
\end{figure}

The ratio $q=  \lambda_{\phi h} / (2 \lambda_\phi)$ determines the behavior of the solution. The momentum modes inside instable regions shown in Fig.~\ref{Lame-fig} have exponential increase in $X_k$. The exponential growth due to the parameteric band structure of the Lam\'e equation is called {\it parameteric resonance}. Namely, it leads to resonance particle production since the occupation number gets amplified accordingly
\begin{equation} \label{nk}
n_k = {\omega_k \over 2} \, \left(  {   \vert  X_k^{\prime} \vert^2   \over \omega_k^2} + \vert  X_k \vert^2   \right) -{1\over 2}    \;,
\end{equation}
where the prime represents the conformal time derivative $a (d/dt)$ and $ \omega_k^2 = k^2  + {\lambda_{\phi h} \over 2 } \Phi_0^2 \;  {\rm cn}^2 \left( x, {1\over \sqrt{2}} \right) $.

However, the linear analysis above breaks down when the interaction term neglected in \eqref{eom} becomes significant and thereby invalidates \eqref{jacobi} and \eqref{Lame}. Besides such a backreaction effect, the generated particle can also rescatter the inflaton background field. Even in the absence of interactions with other fields, self--interaction of the inflaton fragments the homogeneous background field by rescattering, as illustrated in Fig.~\ref{Lame-fig}. To take into account the non--linear and non--perturbative effects, one has to resort to lattice simulations. We have obtained the results with LATTICEEASY \cite{Felder:2000hq}.

  \begin{figure}[t] 
\centering{
\includegraphics[scale=0.32]{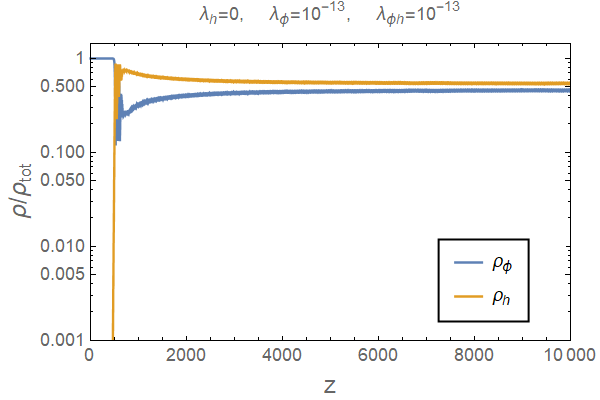}
\includegraphics[scale=0.32]{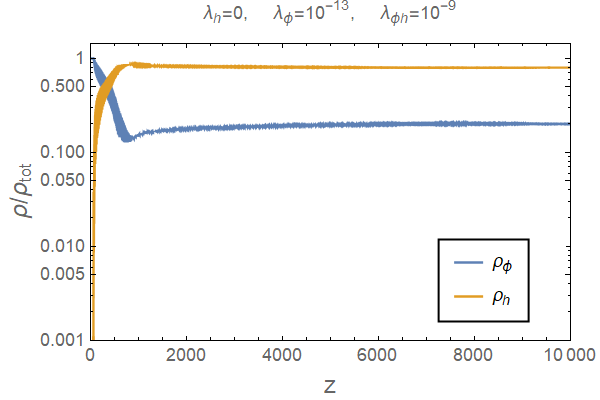}\\
}
\caption{ \label{rho-fig}
The energy fraction of 4 Higgs  d.o.f. and the inflaton as a function of the conformal time $z$,  
  $dz =\sqrt{\lambda_\phi} \,\Phi_0  \,dt/a(t) $ and  $\Phi_0 =1.7$ in Planck units. 
}
\end{figure}

Motivated by the simulation results presented in Fig.~\ref{rho-fig}, we expect the Higgs field to attain the quasi--equilibrium with the SM fields, where the energy is distributed almost democratically among the SM relativistic degrees of freedom,
 \begin{equation}
 {\rho_\phi \over \rho_{\rm tot}} \sim {1\over \# \, {\rm d.o.f.}}
 \end{equation}
When the preheating stage is over, the total number of the inflaton dark matter quanta remains approximately constant.
This allows us to set the $lower$ bound on the dark matter abundance $Y$, which is also invariant from this point.
 The relic abundance $Y$ is defined by $n_\phi/s_{\rm SM}$, where  $n_\phi$ is the inflaton number density and  $s_{\rm SM}$ is the entropy density of the SM fields.
$s_{\rm SM}$  is close to the number density of the SM quanta $n_{\rm SM}$, up to a factor of a few. For 
$107$ SM d.o.f. at high temperature, we find 
\begin{eqnarray}
{\rho_\phi \over \rho_{\rm tot}} \sim {n_\phi \over n_{\rm SM}} \sim {Y \over (n_{\rm SM}/s_{\rm SM})} \gtrsim {1\over 107} \;, \\
Y \gtrsim 10^{-3} \;.
\label{Y}
\end{eqnarray}
This number is far above the observed DM abundance $Y_{\rm obs} = 4.4 \times 10^{-10} \; {\rm GeV}/m_\phi $ for $m_\phi \gtrsim
10$ keV required by the structure formation constraints. Therefore, the resulting Universe is too dark.

\section{Thermal inflaton dark matter}

If $\lambda_{\phi h}$ is sufficiently large, the system may reach thermal equilibrium via processes like $\phi h \rightarrow \phi h, \phi \phi \rightarrow hh$, etc. Let us consider constraints away from the narrow resonance region $m_\phi \simeq 62$ GeV for efficient DM annihilation $\phi \phi \rightarrow \;$SM combined with the XENON1T   bound \cite{Athron:2018ipf}. Then, one reads off
\begin{equation}
\lambda_{\phi h} (1\,{\rm TeV}) \gtrsim 0.25 \;.
\end{equation}
This bound applies at the TeV scale. The renormalization group running generates the inflaton self--coupling at least of the size  $\lambda_{\phi h}^2 /(8\pi^2)$ (ignoring 
a large $log$),
\begin{equation}
\lambda_{\phi } (H) \gtrsim 10^{-3} \;.
\end{equation}
Clearly, the generated coupling $\lambda_\phi$ is inconsistent with the unitarity bound \eqref{uni-bound}.

One may evade the above conclusion by assuming the Higgs resonance, 
\begin{equation}
m_\phi \simeq m_{h_0}/2 \;,
\end{equation}
where $m_{h_0} =125$ GeV is the physical Higgs mass.
In this case, resonant DM annihilation $\phi\phi \rightarrow h \rightarrow {\rm SM}$
 is efficient even for small couplings $\lambda_{\phi h} \gtrsim 10^{-4}$ \cite{Athron:2018ipf}. The correction to the inflaton self--coupling is insignificant and all of the constraints can be avoided. 
 We note however that $| m_\phi - m_h/2 |$ must be below a few GeV, which is not impossible yet rather unnatural.

\section{Conclusion}

The inflaton dark matter model is an economical and intriguing possibility that a single scalar field is responsible for both inflation and dark matter. Based on minimalism, we have examined thermal and non--thermal production of the inflaton dark matter quanta. It is important to take into account non--linear and non--perturbative effects in analyzing non--thermal production of particles during preheating, for which we have performed lattice simulations. We find that at weak couplings only a small fraction of the inflaton energy is converted into the SM radiation, while at large couplings the system reaches quasi--equilibrium. In both cases, dark matter is overabundant. Thermal inflaton dark matter is viable but only within the narrow range of the inflaton mass, such that the inflaton quanta annihilate through the Higgs resonance. The conclusion of the inflaton dark matter model may change in less minimalistic set--ups.


\begin{thebibliography}{99}



\bibitem{Liddle:2008bm}
A.~R.~Liddle, C.~Pahud and L.~A.~Urena-Lopez,
Phys. Rev. D \textbf{77}, 121301 (2008).



\bibitem{Chernikov:1968zm}
N.~A.~Chernikov and E.~A.~Tagirov,
Ann. Inst. H. Poincare Phys. Theor. A \textbf{9}, 109 (1968).


\bibitem{Lerner:2009xg}
R.~N.~Lerner and J.~McDonald,
Phys. Rev. D \textbf{80}, 123507 (2009)


\bibitem{Kofman:1994rk}
L.~Kofman, A.~D.~Linde and A.~A.~Starobinsky,
Phys. Rev. Lett. \textbf{73}, 3195-3198 (1994).


\bibitem{Kofman:1997yn}
L.~Kofman, A.~D.~Linde and A.~A.~Starobinsky,
Phys. Rev. D \textbf{56}, 3258-3295 (1997).

 

\bibitem{Greene:1997fu}
P.~B.~Greene, L.~Kofman, A.~D.~Linde and A.~A.~Starobinsky,
Phys. Rev. D \textbf{56}, 6175-6192 (1997).


 
\bibitem{Khlebnikov:1996wr}
S.~Y.~Khlebnikov and I.~I.~Tkachev,
Phys. Lett. B \textbf{390}, 80-86 (1997).
 
 
 
 \bibitem{Prokopec:1996rr}
T.~Prokopec and T.~G.~Roos,
Phys. Rev. D \textbf{55}, 3768-3775 (1997).


\bibitem{Lebedev:2021zdh}
O.~Lebedev and J.~H.~Yoon,
Phys. Lett. B \textbf{821}, 136614 (2021)


\bibitem{Lebedev:2011aq}
O.~Lebedev and H.~M.~Lee,
Eur. Phys. J. C \textbf{71}, 1821 (2011). 

\bibitem{Lebedev:2021xey}
O.~Lebedev,
Prog. Part. Nucl. Phys. \textbf{120}, 103881 (2021)


\bibitem{Akrami:2018odb}
Y.~Akrami \textit{et al.} [Planck],
Astron. Astrophys. \textbf{641}, A10 (2020).

\bibitem{Bezrukov:2007ep}
F.~L.~Bezrukov and M.~Shaposhnikov,
Phys. Lett. B \textbf{659}, 703-706 (2008)


\bibitem{Bezrukov:2010jz}
F.~Bezrukov, A.~Magnin, M.~Shaposhnikov and S.~Sibiryakov,
JHEP \textbf{01}, 016 (2011)


\bibitem{Felder:2000hq}
G.~N.~Felder and I.~Tkachev,
Comput. Phys. Commun. \textbf{178}, 929-932 (2008).




\bibitem{Athron:2018ipf}
P.~Athron, J.~M.~Cornell, F.~Kahlhoefer, J.~Mckay, P.~Scott and S.~Wild,
Eur. Phys. J. C \textbf{78}, no.10, 830 (2018).



 
 
 

\end{thebibliography}
\end{document}